\documentclass{andp2012}
\usepackage[english]{babel}
\category{Original Paper}
\keywords{Variation of fundamental constants, Atomic and molecular
spectroscopy}
\title{Sensitivity coefficients to variation of fundamental constants}
\author[M.\,G. Kozlov]{Mikhail G. Kozlov\inst{1,2}
\footnote{Corresponding author\quad
E-mail:~\textsf{mihailgkozlov@gmail.com}}}
\author[D. Budker]{Dmitry Budker\inst{3,4,5}}
\address[1]{Petersburg Nuclear Physics Institute of NRC ``Kurchatov
Institute'', Gatchina 188300, Russia}
\address[2]{St.~Petersburg Electrotechnical
University LETI, Prof.~Popov Str.~5, 197376 St.~Petersburg, Russia}
\address[3]{Helmholtz Institute Mainz, Johannes Gutenberg
University, 55099 Mainz, Germany}
\address[4]{Department of Physics, University of
California at Berkeley, Berkeley, California 94720-7300, USA}
\address[5]{Nuclear Science Division,
Lawrence Berkeley National Laboratory, Berkeley, California 94720,
USA}
\shortauthors{M.\,G. Kozlov and D. Budker}
\begin{abstract}
Atoms and molecules can serve as sensitive probes of a possible variation of the fine structure constant $\alpha$ and electron-to-proton mass ratio $\mu$.  Two types of sensitivity coefficients are often used to quantify and compare the sensitivity of different species to the variation of fundamental constants. The dimensionless coefficients $K$ are related to the fractional sensitivity, while dimensional factors $q$ are related to the absolute sensitivity. Here we discuss several common errors and misconceptions regarding these coefficients that frequently appear in the literature.  
\end{abstract}
\shortabstract
\begin{document}
\maketitle


Transition frequencies in atoms and molecules depend on the values of the fundamental constants (FC). The energies of all stationary states of these systems are proportional to the atomic energy unit 
$E_h=4\pi\hbar c R_\infty=m_e e^4/\hbar^2$ 
and additionally depend on the fine structure constant $\alpha=e^2/(\hbar c)$ and the electron-to-proton mass ratio $\mu=m_e/m_p$. Here $m_e$ and $m_p$ are the electron and proton masses, $e$ is the elementary-charge magnitude, $\hbar$ is Plank's constant, $R_\infty$ is Rydberg constant, and $c$ is the speed of light. In principle, all energy levels depend on the nuclear radii (isotope field shifts), but this dependence is usually weak. The hyperfine energy additionally depends on the nuclear moments, most importantly, on the $g$ factors and quadrupole moments of the nuclei. All nuclear parameters, in turn, depend on the QCD energy scale $\Lambda_\mathrm{QCD}$ and on quark masses.

When we search for a possible temporal variation of FC we are studying time-variation of atomic (molecular) frequencies. It is difficult to imagine how to compare directly the frequency today with itself yesterday. Using additional devices, like resonators, or delay lines, introduces time-evolution of the frequency caused, for example, by a possible change of the resonator size, or shape. Instead, we need to measure the ratio of two frequencies and look how this ratio varies in time. This ratio changes only when two frequencies have different dependences on FC. Clearly, the dependence on the atomic energy unit cancels out. Because of that, it is convenient to use atomic units when discussing the sensitivity to a variation of FC.  

The dependence of the atomic optical transitions between different electronic configurations on the mass ratio $\mu$ and on the nuclear radii and moments is weak. We conclude that frequency ratios for atomic optical transitions depend mainly on $\alpha$. This dependence appears due to the relativistic corrections to the energy. Such corrections scale as $\alpha^2 Z^2\times\mathrm{IP}$, where $Ze$ is nuclear charge and IP is the ionization potential. For light atoms $Z\approx 1$ and the corrections are small. For heavy atoms, where $Z\gg 1$, and/or for highly charged ions, where $\mathrm{IP}\gg E_h$, the relativistic corrections become large. Consequently, $\alpha$-dependence of the optical transitions is weak for light atoms and strong for heavy atoms and highly charged ions. 

For molecules, due to the vibrational and rotational energy-level structure, the transition frequencies generally depend on both $\alpha$ and $\mu$. Because of the interplay of different contributions to the molecular energy, the spectra of molecules are much richer and occasionally their dependence on FC may be significantly enhanced.  

\subsection*{Dimensional \& dimensionless sensitivity coefficients}

In 1999 Dzuba, Flambaum, and Webb \cite{DFW99b,DFW99a} introduced $q$ factors, which define how atomic energies depend on $\alpha$:
\begin{equation}
 \delta E = 2q_\alpha\frac{\delta \alpha}{\alpha_0},
 \label{eq:q}
\end{equation}
where $\alpha_0$ is the present value of the fine structure constant. The $q$ factor has dimension of energy. A transition frequency $\hbar\omega_{i,f}=E_f-E_i$ depends on $\alpha$ as:
\begin{equation}
 {\hbar}\delta \omega_{i,f} 
 = 2\left(q_{\alpha}^f-q_{\alpha}^i\right)\frac{\delta \alpha}{\alpha_0}
 \equiv 2q_{\alpha}^{i,f}\frac{\delta \alpha}{\alpha_0}
 \,.
 \label{eq:q_if}
\end{equation}
It is often convenient to introduce a dimensionless sensitivity coefficient $K_\alpha$ for an $i \to f$ transition:
\begin{equation}
 \frac{\delta \omega}{\omega_0} =  K_\alpha  \frac{\delta \alpha}{\alpha_0}\,,
 \qquad
 K_\alpha=\frac{2q_\alpha}{\omega_0}\,,
 \label{eq:K}
\end{equation}
where we skip the indexes $i,f$ for brevity. In order to find $K_\alpha$,  we first calculate the parameters $q_\alpha$ for both levels and then insert $q_\alpha=q_\alpha^f-q_\alpha^i$ in the second equation of \eqref{eq:K}.

Similarly to Eqs.\ (\ref{eq:q},\ref{eq:K}) we can define the sensitivity coefficients to $\mu$-variation:
\begin{align}
 \delta E = q_\mu\frac{\delta \mu}{\mu_0},
 \label{eq:q_mu}
 \qquad
 \frac{\delta \omega}{\omega_0} =  K_\mu  \frac{\delta \mu}{\mu_0}\,,
 \qquad
 K_\mu=\frac{q_\mu}{\omega_0}\,.
\end{align}
Note, that there is an extra factor of 2 in Eq.\ \eqref{eq:q} compared to the first equation in \eqref{eq:q_mu}, which was introduced in \cite{DFW99b,DFW99a} because atomic energies depend on $\alpha^2$.

Let us consider the relevance of the coefficients $q$ and $K$ for various experiments. According to Eqs. (\ref{eq:q},\ref{eq:q_mu}) the factors $q$ give the absolute energy (frequency) shifts for a given change of the FC. On the other hand, the factors $K$ determine the relative frequency change. Therefore, the latter are important when the relative accuracy of the frequency measurement is fixed. For example, this may be the case, when the lines are Doppler broadened. Then the linewidth is proportional to the transition frequency,  $\Gamma_D/\omega = \Delta v/c$, and shifts down to a certain fraction of the linewidth can be experimentally resolved ($\Delta v$ is the width of velocity distribution). This situation is typical for high-redshift astrophysical observations. In high-precision laboratory measurements, on the contrary, Doppler-free spectroscopy and optical combs are often used. Then the experiments are characterized by the absolute accuracy of the frequency measurements. In this case, the $q$ factors are more relevant. 

If there is an accidental degeneracy of two energy levels $i$ and $f$ with different $q$ factors, then according to Eq.\ \eqref{eq:K} the transition sensitivity coefficient $K^{i,f}$ is inversely proportional to the frequency $\omega_{i,f}$ and is strongly enhanced.\footnote{This is not generally true for non-accidental degeneracy. Consider, for example, the fine structure splitting. In atomic units it scales as $\alpha^2 Z^2$ and the sensitivity coefficients are $K_\mathrm{fs}=2$ independently of the size of the splitting.} However, an actual experimental sensitivity to the FC variation is enhanced \textit{only if} the transition frequency $\omega_{i,f}$ is directly measured with high relative accuracy.
The sensitivity coefficients $K$ are \textit{irrelevant} for \textit{indirect} measurements, when small frequency $\omega$ appears as a difference of two measured frequencies,  $\omega=\omega_a- \omega_b$. Here the absolute experimental error does not depend on $\omega$ and the relative experimental error is inversely proportional to $\omega$. This counterbalances growth of the sensitivity coefficient $K$ \cite{NBL04}. This is one of the misunderstandings, which may be found in the literature.

\subsection*{Dependence on the units and conventions}

The factors $q$ have dimension of energy and clearly depend on the choice of units. In atomic physics, some calculations are done using atomic units while others use relativistic units, where the energy unit is $E_r=m_e c^2$ and $E_h=\alpha^2 E_r$. Consequently, the calculated $q$ factors are different depending on the units used. 

Moreover, even if we use the same units, the result depends on the choice of the reference energy. For example, for many-electron atoms the energy of a level $i$ can be defined as the core and valence parts, $E_i= E_\mathrm{core}+\tilde{E}_i$, and the core energy $E_\mathrm{core}$ is often dropped. The core energy is typically much larger than the valence energy $\tilde{E}_i$ and, therefore, $|q_i|\gg |\tilde{q}_i|$. Different definitions of the core may lead to very different values of $\tilde{q}_i$. Of course, when we calculate $q$ factors for the transitions, the core contribution cancels out and transition coefficients $q^{i,f}=\tilde{q}^{i,f}$ depend only on the choice of units. In practice, the $q$ factors are usually calculated with respect to the ground state, thus $q^{i}=q^{0,i}$. 

The dependence of the $q$ factors on units is usually recognized. However, it is often assumed that the dimensionless coefficients $K$ are independent of units. This, in fact, is not the case. According to Eq.\ \eqref{eq:K}, 
\begin{equation}
 K_\alpha = \frac{\partial \omega}{\partial \alpha} \frac{\alpha}{\omega}\,.
 \label{eq:K_der}
\end{equation}
If $\omega$ is in atomic units, then in relativistic units the frequency $\tilde{\omega}$ is 
\begin{equation}
\tilde{\omega}=\alpha^2 \omega\,, 
\end{equation}
and respective sensitivity coefficient is equal to
\begin{equation}
\tilde{K}_\alpha =K_\alpha+2\,.
\end{equation}

As pointed out above, in experiments, we always measure the ratio of two frequencies. For example, in an absolute measurement, we determine the ratio of an atomic frequency to the frequency of the cesium frequency standard. The variation of the frequency ratio due to a variation of $\alpha$ is:  
\begin{equation}
 \frac{\delta\left(\omega_m/\omega_n\right)}{\omega_m/\omega_n}
 =\left(K_\alpha^m-K_\alpha^n\right)
 \frac{\delta\alpha}{\alpha}\,.
 \label{eq:K_ratio}
\end{equation}
We see that the variation of the ratio of frequencies depends on the difference of the sensitivity coefficients and does not depend on the choice of the units. 

We conclude, that the values of the sensitivity coefficients $q$ and $K$ depend on the conventions used. However, the differences in the sensitivity coefficients $K$ are independent of the choice of the system of units and the reference frequency. Therefore, it is always necessary to look at the differences in the sensitivity coefficients of the levels and transitions.

\subsection*{Lattice constant}

Up to now we focused on the energies and frequencies. This is because frequency measurements are by far the most accurate measurements in physics. However, in the literature there are discussions of a possible variation of the lattice constant of crystals \cite{StaFla15,StaFla16a}. The dependence of the lattice constant on the FC can be tested in frequency measurements if one uses a crystal to make an optical cavity.\footnote{Amorphous materials are expected to depend on $\alpha$ in the same way as crystals.} The frequency of the $N^\mathrm{-th}$ longitudinal mode of the cavity of the length $L_c$ is: $\omega^N_c= \pi N c/L_c$. The sensitivity coefficient $K_\alpha^c$ of this  frequency to $\alpha$-variation is linked to the sensitivity coefficient $K_\alpha^a$ of the lattice constant $a$: 
\begin{equation}
 \frac{\delta L_c}{L_c} 
 =\frac{\delta a}{a}
 =K_\alpha^a \frac{\delta\alpha}{\alpha}\,.
 \label{eq:K_a}
\end{equation}
In atomic units $c=\alpha^{-1}$ and therefore,
\begin{align}
 &\frac{d\omega^N_c}{d\alpha}
 =\frac{d}{d\alpha}\left(\frac{\pi N}{\alpha L_c}\right)
 =-\omega^N_c
 \left(\frac{1}{\alpha}+\frac{1}{L_c}\frac{dL_c}{d\alpha}\right)\,,
 \quad\Rightarrow
 \nonumber\\
 &K_\alpha^c=-\left(1 + K_\alpha^a\right)\,.
 \label{eq:K_c}
\end{align}
For crystals made of light elements $|K_\alpha^a|\ll 1$ and $K_\alpha^c\approx -1$. The ratio of some atomic frequency $\omega_\mathrm{at}$ to the cavity frequency has sensitivity to $\alpha$-variation, which is given by Eq.\ \eqref{eq:K_ratio};
\begin{equation}
 \frac{\delta\left(\omega_\mathrm{at}/\omega^N_c\right)}
 {\omega_\mathrm{at}/\omega^N_c}
 \approx\left(K_\alpha^\mathrm{at}+1\right)
 \frac{\delta\alpha}{\alpha}\,.
 \label{eq:K_ratio1}
\end{equation}
This is particularly relevant in view of the recent high-precision searches for alpha variation involving just such a comparison \cite{WMBC16,KOBK18}.

\subsection*{Acknowledgement}

This work was inspired by discussions with A. Borschevsky, J. Ye, Y. Stadnik, and V. Flambaum. 


\begin{thebibliography}{[1]}

\bibitem{DFW99b}
 \textsc{V.\,A. Dzuba},  \textsc{V.\,V. Flambaum},  and  \textsc{J.\,K. Webb}
  \jr{Phys. Rev. Lett.} \textbf{82}, 888 (1999).


\bibitem{DFW99a}
 \textsc{V.\,A. Dzuba},  \textsc{V.\,V. Flambaum},  and  \textsc{J.\,K. Webb}
  \jr{Phys. Rev. A} \textbf{59}, 230 (1999).


\bibitem{NBL04}
 \textsc{A.\,T. {Nguyen}},  \textsc{D.~{Budker}},  \textsc{S.\,K. {Lamoreaux}},
   and  \textsc{J.\,R. {Torgerson}} \jr{Phys. Rev. A} \textbf{69}, 022105
  (2004).


\bibitem{StaFla15}
 \textsc{Y.\,V. {Stadnik}} and  \textsc{V.\,V. {Flambaum}} \jr{Physical Review
  Letters} \textbf{114}, 161301 (2015).


\bibitem{StaFla16a}
 \textsc{Y.\,V. {Stadnik}} and  \textsc{V.\,V. {Flambaum}} \jr{Phys. Rev. A}
  \textbf{93}, 063630 (2016).


\bibitem{WMBC16}
 \textsc{P.~{Wcis{\l}o}},  \textsc{P.~{Morzy{\'n}ski}},  \textsc{M.~{Bober}},
  \textsc{A.~{Cygan}},  \textsc{D.~{Lisak}},  \textsc{R.~{Ciury{\l}o}},  and
  \textsc{M.~{Zawada}} \jr{Nature Astronomy} \textbf{1}, 0009 (2016).


\bibitem{KOBK18}
 \textsc{C.~Kennedy},  \textsc{E.~Oelker},  \textsc{T.~Bothwell},
  \textsc{D.~Kedar},  \textsc{L.~Sonderhouse},  \textsc{E.~Marti},
  \textsc{S.~Bromley},  \textsc{J.~Robinson},  and  \textsc{J.~Ye} \jr{Bulletin
  of the American Physical Society} p.\,H06.00005 (2018).


\end{thebibliography}
\providecommand{\WileyBibTextsc}{}
\let\textsc\WileyBibTextsc
\providecommand{\othercit}{}
\providecommand{\jr}[1]{#1}
\providecommand{\etal}{~et~al.}

\end{document}